\documentclass[10pt,conference]{IEEEtran}
\IEEEoverridecommandlockouts
\usepackage{cite}
\usepackage{amsmath,amssymb,amsfonts}
\usepackage{algorithmic}
\usepackage{graphicx}
\usepackage{textcomp}
\usepackage{xcolor}
\usepackage{booktabs}
\usepackage{url}
\usepackage{todonotes}
\usepackage{color}

\usepackage[implicit=false]{hyperref}
\def\BibTeX{{\rm B\kern-.05em{\sc i\kern-.025em b}\kern-.08em
    T\kern-.1667em\lower.7ex\hbox{E}\kern-.125emX}}
    
\begin{document}

\title{MeetDurian: A Gameful Mobile App \\ to Prevent COVID-19 Infection}

\author{\IEEEauthorblockN{Dongliang Chen}
\IEEEauthorblockA{\textit{Qingdao University}\\
Qingdao, China \\
936418030@qq.com}
\and
\IEEEauthorblockN{Antonio Bucchiarone}
\IEEEauthorblockA{\textit{Fondazione Bruno Kessler (FBK)} \\
\textit{Via Sommarive 18, 38123}\\
Trento, Italy \\
bucchiarone@fbk.eu}
\and
\IEEEauthorblockN{Zhihan Lv}
\IEEEauthorblockA{\textit{Qingdao University}\\
Qingdao, China \\
lvzhihan@gmail.com}
}

\maketitle
\begin{figure*}[h] 
	\centering
	\includegraphics[width=14cm]{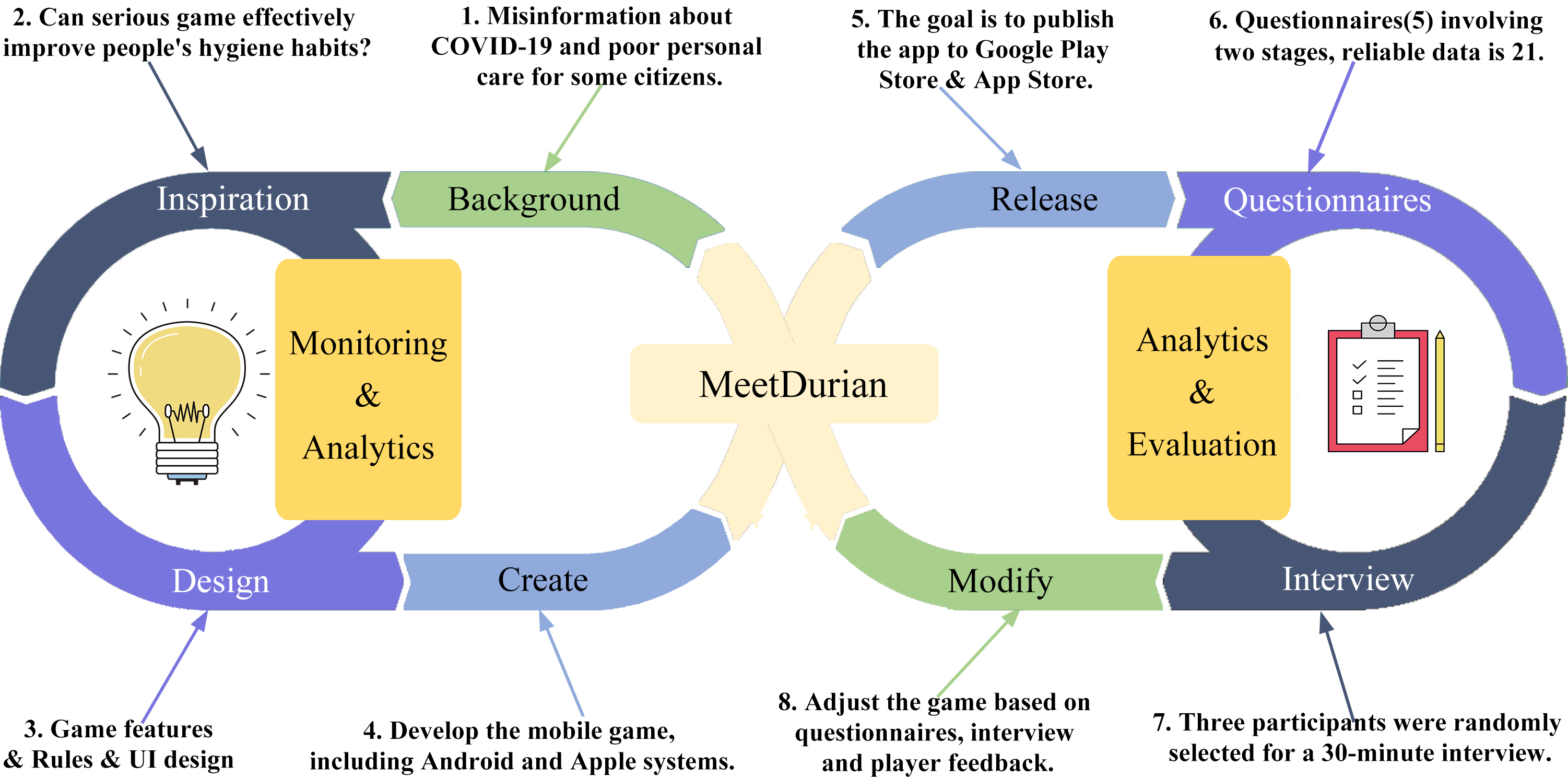}
	\caption{Adapted DSPM for our research.}
	\label{fig:DSPM}
\end{figure*}
\begin{abstract}
The COVID-19 problem has not gone away with the passing of the seasons. Even though most countries have achieved remarkable results in fighting against epidemic diseases and preventing and controlling viruses, the general public is still far from understanding the new crown virus and lacks imagination on its transmission law. In this paper, we propose \textit{MeetDurian}: a cross-platform mobile application that exploits a location-based game to improve users' hygiene habits and reduce virus dispersal. We present its main features, its architecture, and its core technologies. Finally, we report a set of experiments that prove the acceptability and usability of MeetDurian. An illustrative demo of the mobile app features is shown in the following video: \url{https://youtu.be/Vqg7nFDQuOU}.

\end{abstract}
\section{Motivations}
Location-Based Game (LBG) refers to mobile application games that are generally installed on a mobile phone that can be conveniently carried by the user and encourage users to move around frequently. The currently most popular LBG \cite{laato2020location} include but are not limited to: \href{https://www.pokemongo.com}{\textbf{Pokémon GO}} (PG); \href{https://orna.guide}{\textbf{Orna: The GPS RPG}} (Orna); \href{https://www.thewalkingdeadourworld.com}{\textbf{The Walking Dead: Our World}} (TWDOW); \href{https://www.ingress.com}{\textbf{Ingress Prime}} (IP); and \href{https://www.jurassicworldalive.com}{\textbf{Jurassic World Alive}} (JWA). Unfortunately, these games’ main aim is to motivate people to go out, explore, and meet other people. However, these related games made some rule changes during the COVID-19 pandemic. For example, PG’s actions to combat COVID-19 include cancelling the Social Community Day event of Lugia Raid Week and Raid Hour, announcing that they want their players to keep their distance while playing. Similarly, IP cancelled some social events, to protect players from going out and contracting COVID-19. Orna on the other hand recommended that players play at home without having to move around. In conclusion, the developers of LBG did not completely change their game rules but only suggested staying at home or cancel some public gatherings and events. Moreover, it is inevitable for players to buy some necessities for life. Hence, face masks are effective tools to protect themselves against disease. 

The study around diversity in providing science-based information on preventing COVID-19 or personal care games is well-developed \cite{gaspar2020mobile}. However, a literature search did not yield an established work on location-based serious games to solve the above problem during the pandemic, where \textbf{serious game} refers to those games whose main content is to teach knowledge, provide professional training and simulation. From the perspective of prevention and control efforts of basic cases, many countries and regions in the world can effectively control this epidemic. For example, in Italy, the method of changing from the lockdown to more relaxed situations had been adopted. However, it is vital for citizens to observe good hygiene habits, especially focusing on developing long-term habits. 

Because the design of the MeetDurian was motivated by the above background problem, our research can be classified as a problem centered initiation, which the main goal is to develop a location-based application designed to improve users’ hygiene habits. The game logic is as follows: players need to enable the Geo-location service and use email \& password to log in, and they are forced to wear face masks before they can start game. Some durians appear randomly in the vicinity. A user can capture durians by (1) players need to walk closer to a durian, (2) successfully pass the question and answer (Q\&A). Then players can get 1 game point (wrong answer = player's HP - 0.5). Where game point can be used to purchase items, redeem 'Hero Title' in the shop, and players can compare their points with other players in the global ranking interface. Most importantly, we emphasize on the spirit of cooperation in game. We chose the extended design scientific process model (DSPM) \cite{lv2015game}, following the scientific design perspective commonly used in software engineering. Figure \ref{fig:DSPM} illustrates the entire development cycle. In summary, the main contributions are as follows:

\begin{itemize}
    \item We have built a \textit{mobile application} MeetDurian\footnote{\url{https://play.google.com/store/apps/details?id=io.game2.meetdurian}} for Android and iOS operating systems. 
    \item We propose the \textit{DurianToRoads algorithm}, which can intelligently transfer the durian from abnormal locations to the nearest road. Moreover, we integrate face detection of the Google Vision features and have implemented it to detect face masks by analyzing facial feature data.
    \item We have done two \textit{technical evaluation tests}. Test 1 showed that virtual durians can appear evenly and scattered around the players within a particular zone. In test 2, face mask recognition feature can adapt to complex shooting environments, which can accurately identify whether the player is wearing a mask.
    \item We conducted a series of evaluation experiments by questionnaire. And results showed that MeetDurian could improve players’ hygiene habits moderately.
\end{itemize}

\section{MeetDurian Implementation}

In the following sections, we describe the software architecture of the MeetDurian App, its features, and system flow.

\subsection{MeetDurian Architecture}

\begin{figure}
	\centering
	\includegraphics[width=.9\linewidth]{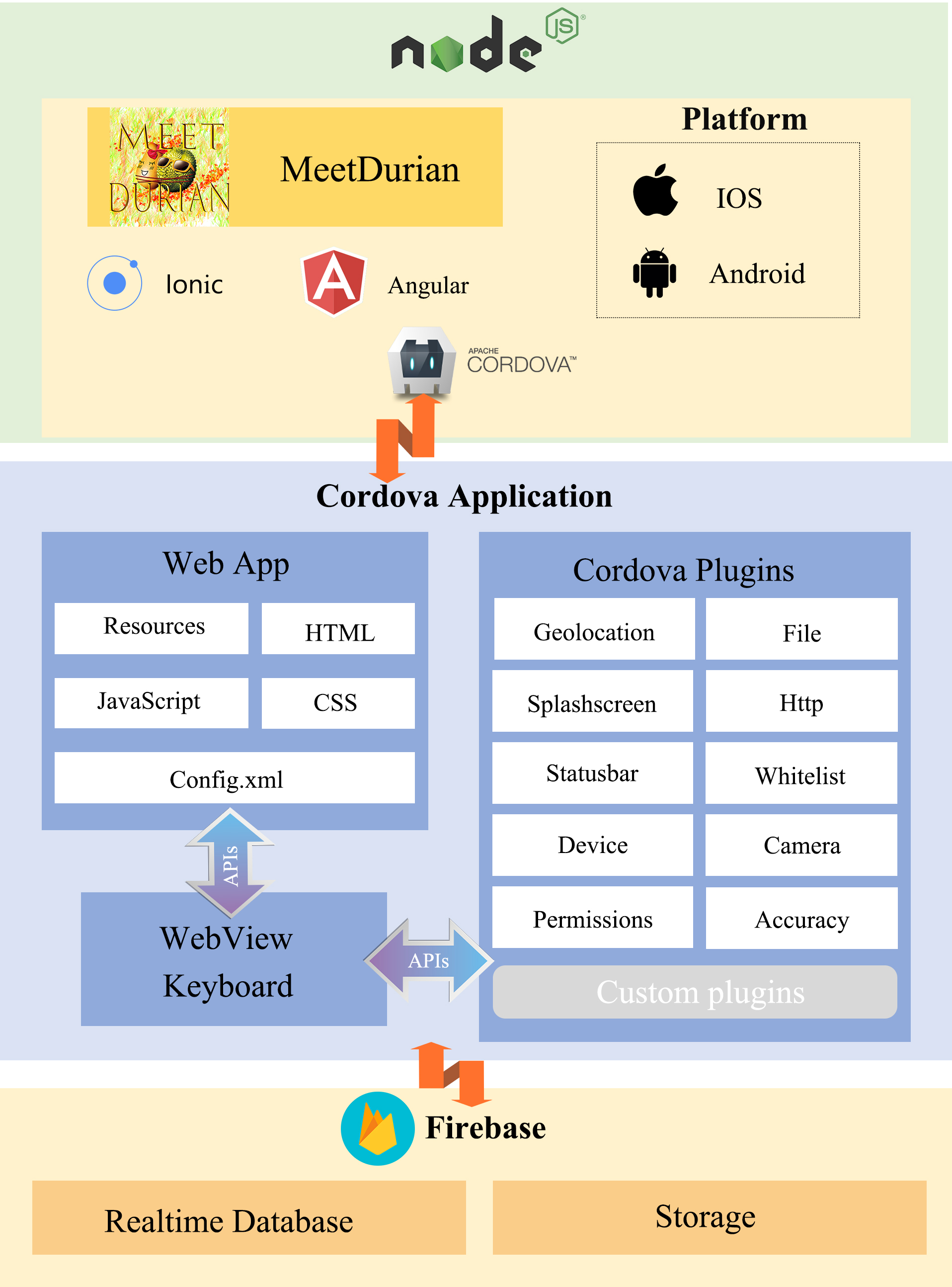}
	\caption{MeetDurian Architecture Overview}
	\label{fig:Architecture}
\end{figure}

Figure \ref{fig:Architecture} shows MeetDurian’s architecture. MeetDurian is an application mainly written in IONIC5+Angular9, based on the node.js environment, and the user interface components were written in HTML, CSS, and JavaScript. The \textbf{Web App} component in Figure \ref{fig:Architecture} is the core code part of the MeetDurian that we have implemented. It includes most of the features, such as the \textit{DurianToRoads algorithm}, and the \textit{Game logic}. 

For the \textit{Face Mask Recognition algorithm}. Google Cloud’s Vision API offers powerful pre-trained machine learning models through Representational State Transfer (REST) and Remote Procedure Call Protocol (RPC) APIs. Consequently, the device interacts with the Google Cloud server through the network to pass the returned data and analyze these data in users' devices for realizing face mask recognition.

\textbf{WebView} can be a component on Android and iOS platforms, which provides the application with its entire user interface. The \textbf{Keyboard} relies on WebView to solve the fuzzy input of the keyboard on the screen and focus elements.

Some of the \textbf{Cordova Plugins} - through some of the the supplied APIs - have been used to: (1) enable the geolocation service without leaving the application, (2) call the front camera of the mobile phone to take photos, (3) obtain the user's latitude and longitude so that durians can appear intelligently. The gray area means that we are not using custom plugins.

Finally, the \textbf{Firebase} component has been used as a back-end database to store global data of the involved players. \textbf{Realtime Database} is hosted on the cloud, the data is stored in JSON format, which synchronized in real-time. Therefore, when users capture durians to gain points, purchase tools in stores to consume points, etc., the data will be automatically updated in real-time, as the same as the leader-boards. \textbf{Storage} is used to store images and videos of MeetDurian. Users can browse pictures and watch videos through their URLs, enabling  players to watch online videos on the novice tutorial.

\begin{figure*}
	\centering
	\includegraphics[width=\linewidth]{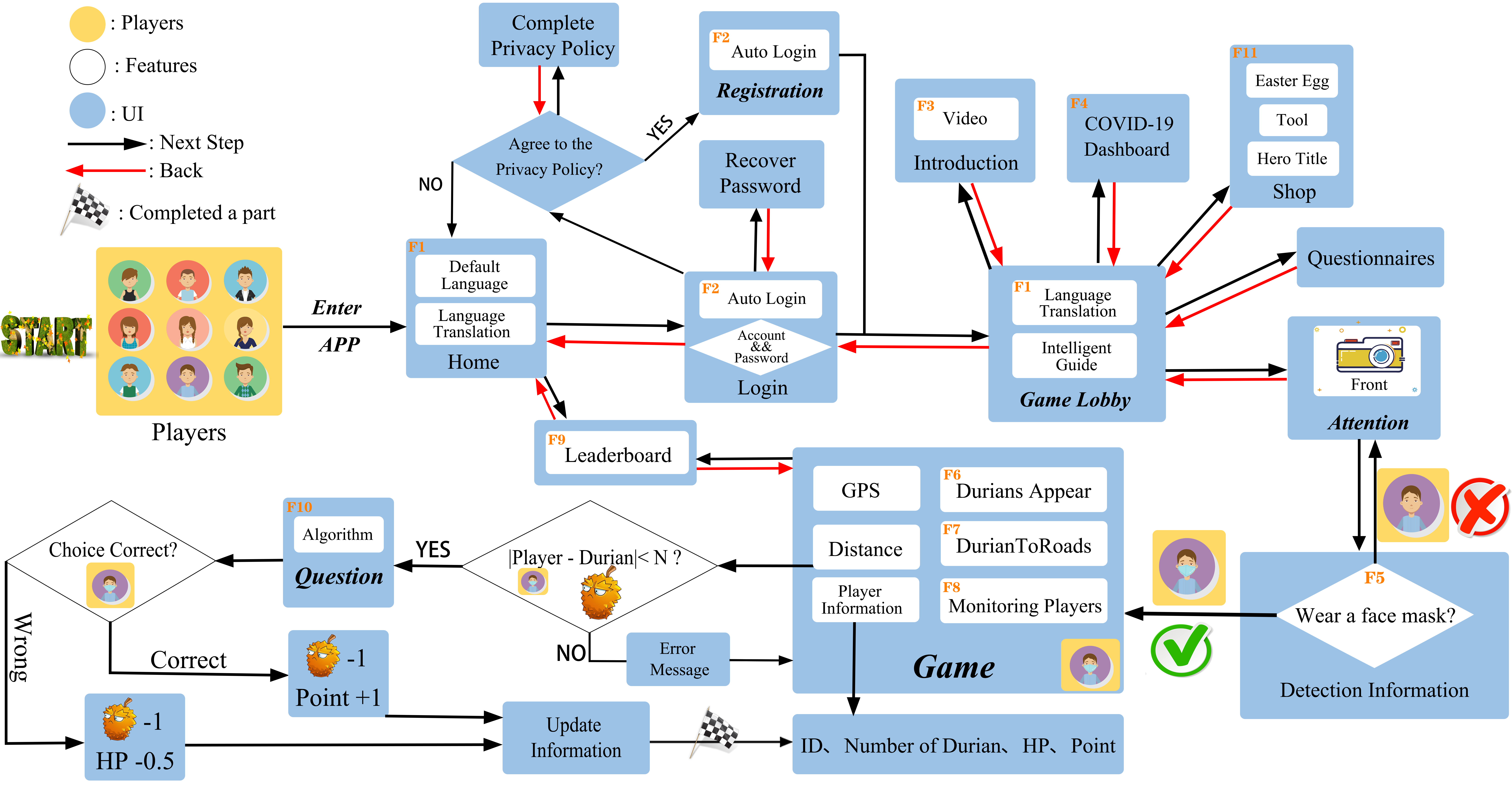}
	\caption{The system flow of MeetDurian. }
	\label{fig:system}
\end{figure*}

\subsection{MeetDurian Features}
The main MeetDurian features are shown in the application flow depicted in Figure \ref{fig:system} and described in the following. 
\begin{itemize}
	
	\item \textbf{F1: Language Selection}. The multi-language version includes Italian, English, and Chinese languages.
	
	\item \textbf{F2: Auto Login}. Automatically jump according to the registered email address and password.
	
	\item \textbf{F3: Introduction}. Novice tutorials (four videos) help players better integrate into the game.

	\item \textbf{F4:  COVID-19 Dashboard}. A dashboard of the COVID-19 data in motion to track the outbreak as it unfolds is constantly updated and shown \cite{dong2020interactive}.
	
	\item \textbf{F5: Face Mask Recognition}. Before entering the game, the mobile phone’s front camera takes pictures to detect the face mask of players. 
	
	\item \textbf{F6: Location Management For Durians}. Six virtual durians will appear randomly around according to their location:
	\begin{itemize}
        \item (1) The location of durians should be randomly and evenly distributed around the player.
		\item (2) The durians’ location should not be too close to the player, and vice-versa.
	\end{itemize}
	
	\item \textbf{F7: DurianToRoads}. Players can re-randomize the abnormal durian in the unreachable area to the nearest road nearby.
	
	\item \textbf{F8: Monitoring players}. Toast message appears at the screen, the green label indicates the player has moved near durians. In contrast, warning the player about the abnormal speed with the red label.
	
	\item \textbf{F9: Real-time Leaderboard}. Players can compare their game scoring (points, levels) with other players in the global ranking page.
	
	\item \textbf{F10: Multiple-choice Question}. The capture of durian needs to be judged based on the player's answer. Moreover, we implemented an algorithm so that when users answer questions, only questions they haven’t answered or for which they gave wrong answers are randomly displayed.
	
	\item \textbf{F11: Shopping}. Including (1) Virtual merchandise and (2) Game Easter egg.
\end{itemize}

\subsection{MeetDurian System Flow}
Exploiting the features described above, we have implemented the MeetDurian game logic as presented in Figure \ref{fig:system}, proceed from left to right and from right to left until in the game interface through face mask detection. MeetDurian mainly contains 16 interfaces. Players can choose multiple languages in the app home page. In the game lobby, new players are asked to watch tutorial videos. In addition, players can also view the COVID-19 dashboard, jump to the shop, fill out questionnaires, take selfies of faces to upload the picture to the server for face mask recognition. In the game, there will be a fixed number of six durians, either success or failure in capturing a durian will count as one part of a game.


\section{EVALUATION}
\begin{figure}
	\centering
	\includegraphics[width=\linewidth]{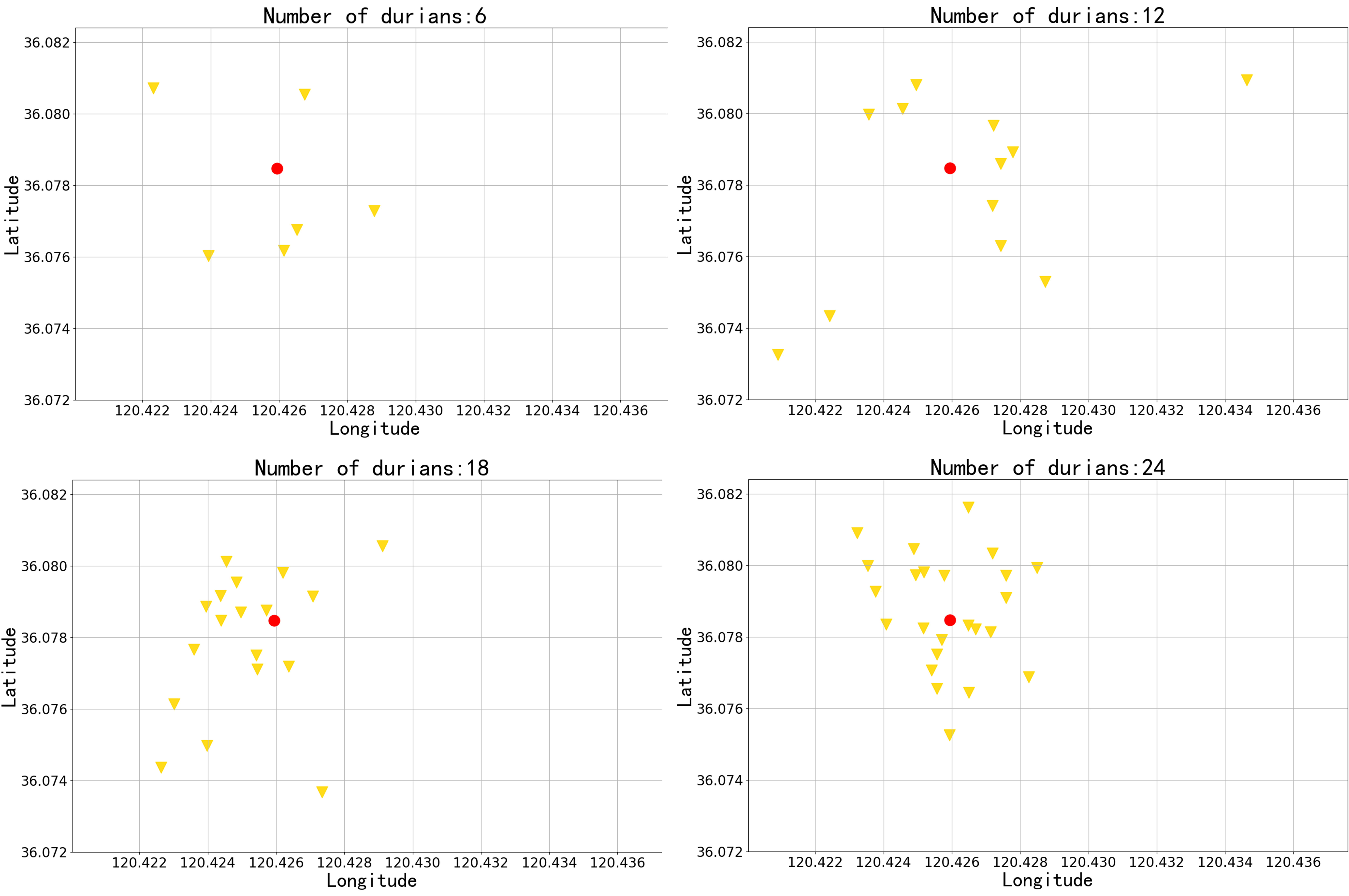}
	\caption{The location distribution of durians (gold) relative to a target point (red).}
	\label{fig:DuriansLocation}
\end{figure}
In this section, we evaluate MeetDurian with the durians' location and face mask recognition, to prove that the acceptability of durians distribution, and the better in real-time and robust of face mask recognition respectively. Finally, a series of questionnaire evaluation results emphasize players can prevent COVID-19 infection with MeetDurian achieving the design purpose of the app.
\subsection{Durians' Location}
We conducted several simulation experiments, using the latitude and longitude of the four corners of a certain university area as the scope, and the latitude and longitude of the university library as the experimental object. Some experimental evaluations of the durians' distribution are shown in Figure \ref{fig:DuriansLocation}. The number of durians increases from left to right. The Golden inverted triangle are the randomly generated location of the virtual durians, and the red dot is the latitude and longitude of the university library.

Although there are some durians outside the target range, the results show that our approach performs much better as the number of durians increases, which demonstrates that the function could better evenly distribute the durians within a specific area while providing a better gaming experience to the gamers. In addition, different tests may correspond to different results. All latitude and longitude data are displayed through a \textbf{GitHub repository}\footnote{\url{https://github.com/Finedl/ScatterPlot}}, along with relevant test code.

\subsection{Face Mask Recognition}
We exploited the face detection method in Google Cloud Vision to perform masked face recognition for this task. In the case of good picture quality, face detection can detect multiple faces in the picture and have high robustness. The main purpose of adopting this method is used to detect the main facial features and can obtain about 27 relevant confidence points for the eyes, nose, mouth, cheeks, etc, in total. We perform data processing and analysis on these large numbers of confidence points to recognize masks’ tasks (i.e., average waiting time \textless 1 second, under good network conditions). In addition, \href{https://firebasestorage.googleapis.com/v0/b/meetdurian.appspot.com/o/Face.pdf?alt=media&token=d1723192-a56d-401c-9423-569bd6a39529}{\textbf{supplementary material}} contains a conceptual figure and tests for complex shooting environments.

\subsection{Questionnaire}
To test the potential of MeetDurian in improving players’ living habits or personal care and raising the awareness of public health and safety awareness, we designed our questionnaires: \href{https://www.wjx.cn/vj/YjAyylv.aspx}{\textbf{Q1, game immersion}}; \href{https://www.wjx.cn/vj/YDarylv.aspx}{\textbf{Q2, user workload evaluation}}; \href{https://www.wjx.cn/vj/m7X0sVI.aspx}{\textbf{Q3, social acceptance}}; \href{https://www.wjx.cn/vj/rggdh9Y.aspx}{\textbf{Q4, learning outcomes}}; \href{https://www.wjx.cn/vj/wFqsUvC.aspx}{\textbf{Q5, personal hygiene}}. \href{https://www.wjx.cn/wxloj/datafullscreen.aspx?activity=104969107}{\textbf{Online visualization data}} shows the results of the questionnaire, where Q5 includes two groups of controlled experiments.  It is obvious that the results are positive, for example, the percentage increases in the number of players bathing and bedroom cleaning are particularly significant, increasing by 52.35\% and 42.82\% respectively. For personal belongings such as mobile phones and keys, 57.14\% of players are willing to use alcohol to keep them clean after the game. And a more detailed breakdown of the results and analysis can be found on the \href{https://firebasestorage.googleapis.com/v0/b/meetdurian.appspot.com/o/Analysis-MeetDurian.pdf?alt=media&token=a48c63dd-e5ae-443f-989e-d075384f9936}{\textbf{expanded worksheet}}.
 
\section{Conclusion and Future Work}
In this paper, we have presented the entire development architecture and related technologies of MeetDurian, which contributed to researchers and developers.

As a future work, we plan to add new game features, including durians can generate in a un-crowded area, setting strict social distance by calculating the distance between different players every second. In addition, COVID-19 will pass from pandemic to prosaic in the future,  to increase engagement among the players, we plan to extend the game logic introducing new gamification mechanics and dynamics. for example, dedicated (single, multiple) players challenges that will also encourage players to behave cooperatively to achieve common goals.

\section{ACKNOWLEDGMENTS}
This work was supported in part by the National Natural Science Foundation of China (NSFC) under Grant Nos. 61902203, Key Research and Development Plan - Major Scientific and Technological Innovation Projects of ShanDong Province (2019JZZY020101).

\bibliographystyle{unsrt}
\bibliography{bib}

\end{document}